\documentclass[journal=achemso-jacsat,manuscript=communication,etalmode=truncate,keywords=true]{achemso}
\setkeys{acs}{articletitle=true,maxauthors=10,etalmode=truncate,keywords=true}
\usepackage{amsmath,amssymb,units,txfonts}
\usepackage{dblfloatfix}
\usepackage{amsfonts}
\usepackage{wrapfig}
\usepackage{helvet}
\usepackage{graphicx}
\usepackage[usenames,dvipsnames]{color}
\usepackage{colortbl,cuted}
\definecolor{JPCCBlue}{RGB}{34,80,169}
\definecolor{JCPCGreen}{RGB}{11,122,64}
\definecolor{StyleColor}{RGB}{34,80,169}
\definecolor{NLRed}{RGB}{215,24,30}
\usepackage[bookmarks=true,colorlinks=true,urlcolor=StyleColor,linkcolor=StyleColor,citecolor=StyleColor]{hyperref} 
\usepackage{multirow}
\usepackage{xr}
\usepackage{lineno}
\usepackage{units,xfrac}
\usepackage{upgreek}
\DeclareMathOperator{\Imag}{Im}
\usepackage{units}
\usepackage{bm}
\newcommand{\brho}{\mathbf{R}}
\newcommand{\bQ}{\mathbf{Q}}
\newcommand{\rv}{\textbf{r}}

\newcommand{\Qv}{\textbf{Q}}
\newcommand{\Xv}{\textbf{X}}
\newcommand{\dv}{{d}}
\newcommand{\EgapQP}{E_{\textit{gap}}^{\text{QP}}}
\newcommand{\Eu}{E$_{1\mathrm{u}}^1$}
\newcommand{\Bu}{B$_{3\mathrm{u}}$}

\definecolor{abstractcolor}{RGB}{255,243,201}
\makeatletter\newenvironment{abstractbox}{%
   \begin{lrbox}{\@tempboxa}\begin{minipage}{0.988\textwidth}}{\end{minipage}\end{lrbox}%
   \colorbox{abstractcolor}{\usebox{\@tempboxa}}
}\makeatother

\renewcommand*\authorfont{\flushleft}
\renewcommand*\affilfont{\mdseries\flushleft\textrm}
\renewcommand*\titlesize{\Large}
\renewcommand*\titlefont{\flushleft\sf\bfseries}
\def\refname{\sffamily\color{StyleColor}\large$\blacksquare$\normalsize\ REFERENCES}
\usepackage{titlesec}
\titleformat{\section}{\bfseries\sffamily\color{StyleColor}}{\thesection.~}{0pt}{}
\titleformat{\subsection}[runin]{\bfseries\sffamily\normalsize}{\indent\thesubsection.~}{0pt}{}[.]
\titlespacing{\subsection}{0pt}{0pt}{*1}
\titleformat{\subsubsection}{\bfseries\sffamily\normalsize}{\thethesubsection.~}{0pt}{}
\titlespacing{\subsubsection}{0pt}{0pt}{*0}

\title{Tailoring a Molecule's Optical Absorbance Using Surface Plasmonics}
\author{Duncan John Mowbray}
\email{duncan.mowbray@gmail.com}
\affiliation[YT]{\footnotemark[2]{\ } School of Physical Sciences and Nanotechnology, Yachay Tech University, Urcuqu\'{\i} 100119, Ecuador}
\alsoaffiliation[DIPC]{\newline\footnotemark[3]{\ } Donostia International Physics Center (DIPC), Paseo de Manuel de Lardizabal 4, ES-20018 San Sebast{\'{i}}an, Spain}
\alsoaffiliation[UPV]{\newline\footnotemark[5]{\ } Nano-Bio Spectroscopy Group and ETSF Scientific Development Center, Departamento de F{\'{\i}}sica de Materiales, Universidad del 
Pa{\'{\i}}s Vasco UPV/EHU, ES-20018 San Sebasti{\'{a}}n, Spain}
\author{Vito Despoja}
\email{vito@phy.hr}
\affiliation[IPZ]{\newline\footnotemark[4]{\ } Institute of Physics, Bijeni\v{c}ka 46, HR-10000, Zagreb, Croatia}
\alsoaffiliation[UZ]{\newline$^\|$Department of Physics, University of Zagreb, Bijeni\v{c}ka 32, HR-10000 Zagreb, Croatia}
\alsoaffiliation[DIPC]{\newline\footnotemark[3]{\ } Donostia International Physics Center (DIPC), Paseo de Manuel de Lardizabal 4, ES-20018 San Sebast{\'{i}}an, Spain}

\begin{document}
\maketitle

\begin{strip}

\noindent{\color{StyleColor}{\rule{\textwidth}{0.5pt}}}
\begin{abstractbox}
  \vspace{0.5em}
  \begin{wrapfigure}[13]{@{}r}{3.25in}
    \vspace{-1em}
    \includegraphics[width=3.25in]{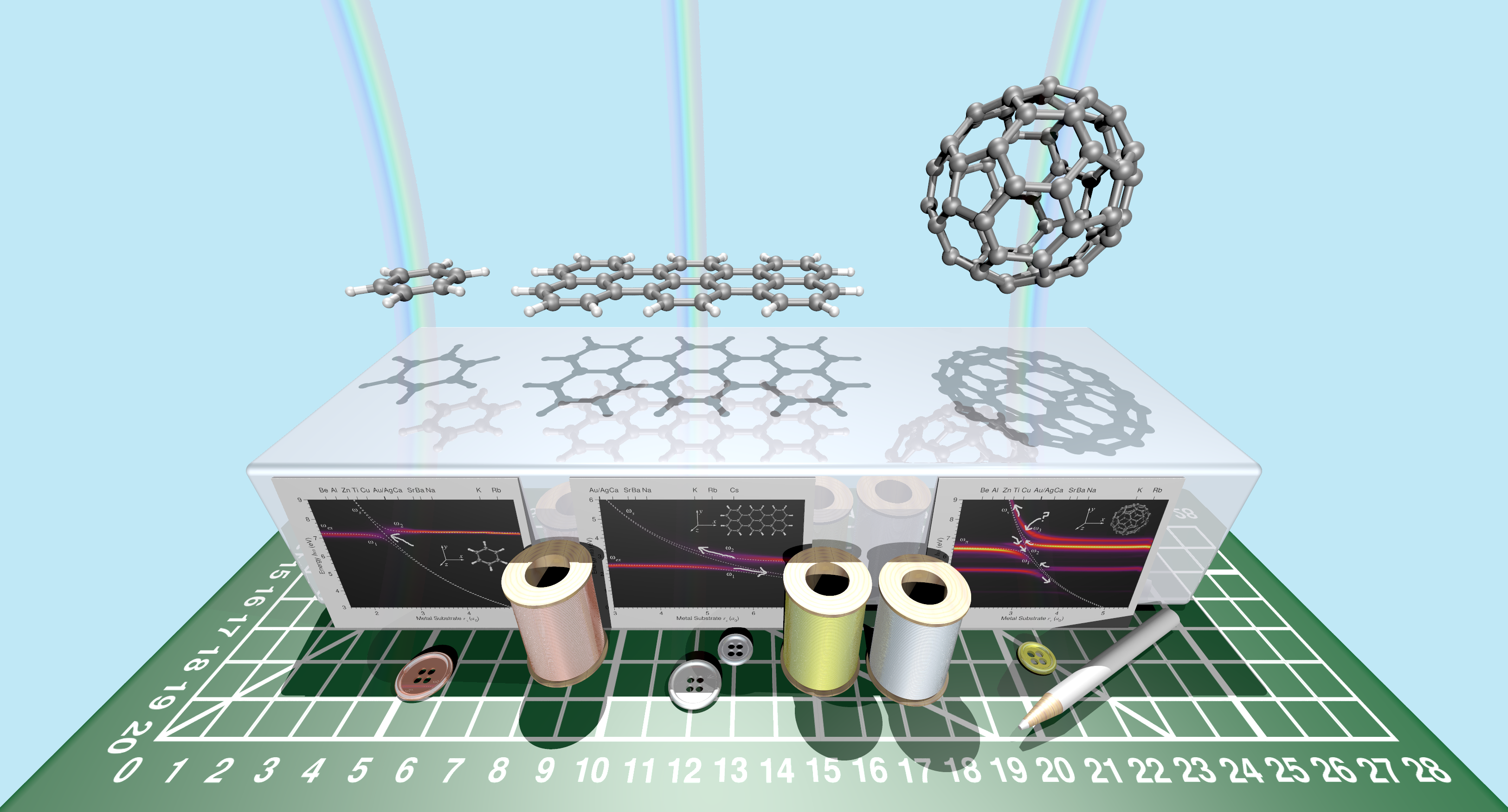}
  \end{wrapfigure}
  \noindent\textbf{\color{StyleColor}{ABSTRACT:}}
  Understanding  the interaction of light with molecules physisorbed on substrates is a fundamental problem in photonics, with applications in biosensing, photovoltaics, photocatalysis, plasmonics, and nanotechnology.  However, the design of novel functional materials \emph{in silico} is severely hampered by the lack of robust and computationally efficient methods for describing both molecular absorbance and screening on substrates.  Here we employ our hybrid $G_0[W_0+\Delta W]$-BSE implementation, which incorporates the substrate via its screening $\Delta W$ at both the quasiparticle $G_0W_0$ level and when solving the Bethe-Salpeter equation (BSE).  We show this method can be used to both efficiently and accurately describe the absorption spectra of physisorbed molecules on metal substrates and thereby tailor the molecule's absorbance by altering the surface plasmon's energy.    Specifically, we investigate how the optical absorption spectra of three prototypical $\pi$-conjugated molecules: benzene (C$_6$H$_6$), terrylene (C$_{30}$H$_{16}$) and fullerene (C$_{60}$), depends on the Wigner-Seitz radius $r_s$ of the metallic substrate.  To gain further understanding of the light--molecule/substrate interaction, we also study the bright exciton's electron and hole densities and their interactions with infrared active vibrational modes.  Our results show that (1) benzene's bright \Eu{} exciton at 7.0~eV, whose energy is insensitive to changes in $r_s$, could be relevant for photocatalytic dehydrogenation and polymerization reactions, (2) terrylene's bright \Bu{} exciton at 2.3~eV hybridizes with the surface plasmon, allowing the tailoring of the excitonic energy and optical activation of a surface plasmon-like exciton, and (3) fullerene's $\pi-\pi^*$ bright and dark excitons at 6.4 and 6.8~eV hybridize with the surface plasmon, resulting in the tailoring of their excitonic energy and the activation of both a surface plasmon-like exciton and a dark quadrupolar mode via symmetry breaking by the substrate. This work demonstrates how a proper description of interfacial light--molecular/substrate interactions  enables the prediction, design, and optimization of technologically relevant phenomena \emph{in silico}.
  \vspace{0.5em}
\end{abstractbox}
\noindent{\color{StyleColor}{\rule{\textwidth}{0.5pt}}}
\end{strip}

\section{INTRODUCTION}

A proper understanding of how light interacts with the  molecule--substrate interface is essential when designing technologically relevant materials for photonic applications\cite{SurfPlasmNature2003,Plasmonics2016}. These include optoelectronics\cite{Optoelectronics2004,Optoelectronics2012,MolecularPlasmonicsACSPhotonics2019}, photocatalysis\cite{FujishimaReview,Henderson2012,TiO2PhotocatalysisChemRev2014,OurJACS}, bio/chemical sensing\cite{biosensors,LabelFreePlasmonicBiosensors2009,LSPRS_ChemRev,SingleMoleculeDetection2012,BiosensingNatPhotonics2012,TerryleneAbsorbance,BiosensingRev2018,PlasmonicBiosensor2018,2DBiosensing2019}, organic photovoltaics (OPVs)\cite{PolymerFullerenePhotovoltaic,Fullerene_apl1,FullereneOnlyPhotovoltaic,PolymerSolarCellsRev,DyeSensitizedSolarCellsRev,MowbrayExcitons2016}, and plasmonics\cite{SurfPlasmNature2003,PlasmonNatPhotonics2016,2DBiosensing2019,Submitted,LightMatterPlasmonics2019,MolecularPlasmonicsACSPhotonics2019}.  For example, molecule--surface coupling can modify both the energies and intensities of the single-particle spectra, electronic excitations, and phonons, thereby affecting the optical absorption of the molecule.  These effects may be exploited to tailor the optoelectronic properties of the molecule via the substrate's screening interactions.
For these reasons, sufficiently accurate and efficient computational methods are required to direct experimental research towards materials with optimal properties for specific technological applications.

Amongst the computational design community, density functional theory (DFT), a ground-state single-particle method,  is the method of choice for the accurate and efficient description of the potential energy landscape\cite{DFTMarques}.  However, by construction, DFT is not applicable to the description of electronic excitations \cite{Rubio}.  To improve DFT's description at the Kohn-Sham (KS) level of the electronic energies $\epsilon_n$, one may (1) include the single-point derivative discontinuity correction from the exchange potential $\Delta_x$ (GLLBSC) \cite{GLLBSC}, (2) incorporate a two-point constant screening ($\varepsilon \sim 4$)\cite{Marques,MiganiLong
} into the exchange and correlation (xc) functional (HSE06\cite{HSE}), (3) replace the xc potential contribution with the self energy $\Sigma(\omega)$ at the two-point quasiparticle (QP) $G_0W_0$ level \cite{GW,Rubio,KresseGW}, which requires the inclusion of many unoccupied bands ($\sim 8$ bands per C atom\cite{Rubio,DuncanGrapheneTDDFTRPA,OurJACS,Bnz_absorption_theory2,Submitted}), or (4) do so self-consistently in the Green's function $G$ (scQP$GW_0$\cite{KresseGW}) and the screening $W$ (scQP$GW$\cite{QSGW,KressescGW} or scQP$GW1$\cite{OurJACS}).  Besides the electronic structure, a proper description of the excited states involved in light--matter interactions requires the (1) single-point oscillator strengths or matrix elements, (2) two-point screening within the random phase approximation (RPA), and (3) four-point excitonic effects such as electron-hole binding from the Bethe-Salpeter equation (BSE)\cite{Rubio}.  Both the four-point nature and the requirement of many bands for $G_0W_0$-BSE makes this most robust method \cite{GWtheory,Rubio,StevenLouie2006,OurJACS,MiganiLong} unfeasible for all but the simplest molecular--substrate interfaces\cite{MowbrayExcitons2016}.

To make calculations of interfacial light--matter interactions feasible, we have developed our own implementation \cite{Bnz_absorption_theory2,Submitted} of the $G_0[W_0+\Delta W]$-BSE method\cite{StevenLouie2006,G0W0DeltaW2010,Spataru,KristianQEHModelNanoLett2015,G0W0DeltaW2017}. In this approach,  the substrate is included only via its screening $\Delta W$ at both the quasiparticle $G_ 0W_0$ and BSE level.  This dramatically reduces the unit cell, number of electrons, and number of unoccupied bands required in the $G_0W_0$ and BSE calculations, since the substrate is only included through its screening $\Delta W$.  As we shall show,  this vast simplification still provides an accurate description of the molecule--substrate interface's optoelectronic properties.

In our previous $G_0[W_0+\Delta W]$-BSE studies \cite{Submitted,Bnz_absorption_theory2} we have concentrated on the height dependence of the molecule--substrate interface's optical absorbance for a given material.  Herein, we propose a systematic method of ``tailoring'' a molecule's optical absorbance using the substrate's Wigner-Seitz radius $r_s$.  This could be accomplished through the introduction of various types of metal or metal-alloy nanoparticles with tailored Wigner-Seitz radii.

In this work, we study the optical absorbance, excitonic densities, and Fourier transform infrared (FTIR) spectra of three representative $\pi$-conjugated organic molecules on metallic surfaces: (1) the smallest building-block benzene (C$_6$H$_6$), (2) the rectangular ortho- and peri-fused polycyclic arene terrylene (C$_{30}$H$_{16}$), and (3) the spherical carbonaceous truncated icosahedral fullerene (C$_{60}$).  In so doing, we consider a wide range of interfacial optoelectronic phenomena, including the photocatalytic dehydrogenation and polymerization of benzene, the tailoring of optical absorbance via surface plasmonics in terrylene, and the ``turning on'' of fullerene's dark excitonic modes through symmetry breaking by the substrate\cite{Submitted}.

\section{METHODOLOGY}

The computational parameters employed in our DFT, quasiparticle $G_0W_0$, BSE, RPA, and $G_0[W_0+\Delta W]$-BSE calculations are summarized in Table~\ref{Table1}.

\noindent{\color{StyleColor}{\rule{\columnwidth}{1pt}}}\\[-1.5em]
\begin{table}[!ht]
    \caption{\textrm{\bf Computational Parameters Employed in \emph{G}$_{\text{0}}$\emph{W}$_{\text{0}}$, Bethe-Salpeter Equation (BSE), Random Phase Approximation (RPA), and \emph{G}$_{\text{0}}$[\emph{W}$_{\text{0}}$+$\Delta$\emph{W}]-BSE Calculations}
  }\label{Table1}
  \begin{tabular}{llccc}
    \multicolumn{5}{>{\columncolor[gray]{0.9}}c@{}}{}\\[-0.5em]
    \multicolumn{1}{>{\columncolor[gray]{.9}}c@{}}{symbol} &
      \multicolumn{1}{>{\columncolor[gray]{.9}}c@{}}{definition} &
      \multicolumn{1}{>{\columncolor[gray]{.9}}l@{}}{benzene} &
      \multicolumn{1}{>{\columncolor[gray]{.9}}l@{}}{terrylene} &
      \multicolumn{1}{>{\columncolor[gray]{.9}}l@{}}{fullerene} \\[0.5mm]
      \\[-0.5em]
      & molecular formula & C$_6$H$_6$ & C$_{30}$H$_{16}$ & C$_{60}$\\
      $L$ & length of unit cell ($a_0$) & 22.50 & 45.69 & 45.69 \\
      $N_e$ &\# of electrons & 30 & 136 & 240\\
      & & \multicolumn{3}{c}{$G_0W_0$ eigenvalues $\tilde{\varepsilon_i}$}\\\cline{3-5}
      $N_{\mathrm{occ}}$ & \# of occupied levels & 15 & 68 & 120\\
      $N_{\mathrm{unocc}}$ & \# of unoccupied levels & 57 & 256 & 456\\
      & & \multicolumn{3}{c}{$G_0W_0$-BSE}\\\cline{3-5}
      $N_{\mathrm{occ}}$ & \# of occupied levels & 15 & 68 & 120\\
      $N_{\mathrm{unocc}}$ & \# of unoccupied levels & 45 & 120 & 120\\
      & & \multicolumn{3}{c}{RPA Screening $W$}\\\cline{3-5}
      $N_{\mathrm{occ}}$ & \# of occupied levels & 15 & 33 & 120\\
      $N_{\mathrm{unocc}}$ & \# of unoccupied levels & 55 & 147 & 180\\
      & & \multicolumn{3}{c}{$G_0[W_0+\Delta W]$-BSE}\\\cline{3-5}
      $N_{\mathrm{occ}}$ & \# of occupied levels & 3 & 1 & 14\\
      $N_{\mathrm{unocc}}$ & \# of unoccupied levels & 3 & 1 & 14\\
\end{tabular}
\noindent{\color{StyleColor}{\rule{\columnwidth}{1pt}}}
\end{table}

\subsection{Isolated Molecules}
Calculations of the isolated molecules have been performed with the plane-wave DFT code \textsc{vasp}\cite{kresse1996b}  within the projector augmented wave (PAW) scheme \cite{kresse1999}, using the local density approximation (LDA) \cite{LDA} for the exchange and correlation (xc) functional.  We model the molecules using a periodically repeated cubic unit cell with $L = 22.50~a_0 \approx 11.91$~\AA{} for benzene, and $L =45.69~a_0 \approx 24.18$~\AA{} for terrylene and fullerene, as shown in Table~\ref{Table1}. Since there is no intermolecular overlap, the ground state electronic density is calculated only at the $\Gamma$ point.  The geometries have been fully relaxed, with all forces $\lesssim$ 0.02 eV/\AA.  We employ a plane-wave energy cutoff of 445 eV, an electronic temperature $k_B T\approx0.2$ eV with all total energies extrapolated to $T\rightarrow 0$ K, and a PAW LDA pseudopotential for carbon.  

To calculate the quasiparticle $G_0W_0$ eigenvalues $\tilde{\varepsilon_i}$ for the isolated molecules, one must include an increased number of unoccupied levels to describe the continuum.  The benzene, terrylene, and fullerene molecules have $30$, $136$, and $240$ valence electrons, corresponding to $15$, $68$, and $120$ doubly-occupied valence levels, respectively.  We included 72, 324, and 576 levels in our $G_0W_0$ calculations for benzene, terrylene, and fullerene, respectively, as given in Table~\ref{Table1}.  This is equivalent to 4.8 unoccupied levels per occupied level or 9.6 levels per C atom and 2.4 levels per H atom. For terrylene and fullerene this proved sufficient to converge the quasiparticle $G_0W_0$ electronic gap $\EgapQP$ within 15~meV, whereas for benzene a fully converged $\EgapQP$ is red shifted by $\approx 0.4$~eV.  This has been taken into account by applying this red shift \emph{a postiori} to our $G_0W_0$-BSE spectrum and exciton energies for gas-phase benzene.

The screening $W$ is obtained from  the  dielectric function, based on the KS wavefunctions \cite{Angel}.  This is calculated using linear response time-dependent DFT in reciprocal space and the frequency domain within RPA including local crystal field effects, i.e., solving Dyson's equation for the fully interacting head of the dielectric function \cite{KresseG0W0}.  To calculate the dielectric function\cite{KresseG0W0} we employed an energy cutoff of 40 eV for the number of \textbf{G}-vectors, and a non-linear sampling of 40 frequency points for the dielectric function up to 200 eV.  This large energy range is necessary to include the main features of fullerene's dielectric response \cite{Scully05,Moskalenko2012}.  From these calculations we obtained converged quasiparticle $G_0W_0$ eigenvalues $\tilde{\varepsilon}_i$ for the isolated fullerene molecule.

We obtain optical absorption spectra for the isolated molecules using the quasiparticle $G_0W_0$ eigenvalues $\tilde{\varepsilon}_i$ within the Bethe-Salpeter equation ($G_0W_0$-BSE)\cite{Rubio,BSE,KresseBSE}.  For solving the BSE, we reduce the number of unoccupied levels included to 45, 120, and 120 for benzene, terrylene, and fullerene, respectively, as shown in Table~\ref{Table1}.  

The $\lambda$th exciton, i.e., the two-point excitonic wavefunction, may be expressed in terms of the Kohn-Sham (KS) wavefunctions $\psi_n$ as
\begin{equation}
  \Psi_\lambda(\rv_e,\rv_h) = \sum_{nm}A_{\lambda}^{nm} \psi_n(\rv_e)\psi_m^*(\rv_h)
\end{equation}
where $A_{\lambda}^{nm}$ are the exciton's coefficients in the KS basis, $n$ and $m$ are the indices of occupied and unoccupied KS levels, and $\rv_e$ and $\rv_h$ are the electron and hole coordinates, respectively.  

The $\lambda$th exciton's electron and hole densities $\rho_{\lambda,e}$ and $\rho_{\lambda,h}$ obtained upon integration with respect to the position of the hole and electron, respectively, are then\cite{MowbrayExcitons2016}
\begin{eqnarray}
  \rho_{\lambda,e}(\rv_{e}) &=& \sum_{nm}\sum_{n'}A_{\lambda}^{nm}A_{\lambda}^{n' m}\psi_n(\rv_e)\psi_{n'}^*(\rv_e),\\
  \rho_{\lambda,h}(\rv_{h}) &=& \sum_{nm}\sum_{m'}A_{\lambda}^{nm}A_{\lambda}^{nm'}\psi_m(\rv_h)\psi_{m'}^*(\rv_h).
\end{eqnarray}
It is important to note that, since the electron and hole densities are effectively ``averaged'' over the hole and electron positions, respectively, they are expected to have no net dipole. For this reason they may be effectively interpreted as an averaging of the electron and hole densities with respect to time.

To probe the exciton coupling to infrared (IR) active vibrational modes of the molecule, we have calculated the FTIR spectra as implemented in the ASE package\cite{ASE0,ASE}.  Here, the infrared intensity for the $\lambda$th vibrational mode is given by\cite{PorezagFTIR96}
\begin{equation}
  I_\lambda^{\textrm{IR}} = \frac{\mathcal{N}\pi}{3c} \left|\sum_{a=1}^{N_a}\nabla_{\rv_a} \dv \cdot \Xv_{\lambda,a}\right|^2,
\end{equation}
where $\mathcal{N}$ is the particle density, $c$ is the speed of light, $\rv_a$ is the coordinate of atom $a$, $N_a$ is the number of atoms, $\dv$ is the molecule's electric dipole moment, and $\Xv_{\lambda,a}$ is the $\lambda$th eigenvector of the dynamical or hessian matrix $\mathcal{H}_{a' a} = \nabla_{\rv_{a'}} \otimes \nabla_{\rv_a} E$, where $E$ is the total energy of the system.  For a more detailed description of this theory we refer the reader to ref~\citenum{SpectraRef}\nocite{SpectraRef}.

\subsection{Molecule--Surface Interactions}
The three $\pi$-conjugated molecules considered herein typically physisorb on metal substrates through long-ranged van der Waals interactions.  This generally results in the molecule oriented parallel to the surface in the $x y$-plane with a minimum molecule--surface separation of $z_0 \approx 6~a_0 \approx 3.18$~\AA. For example, Perdew--Burke--Ernzerhof van der Waals (PBE+vdW) calculations of benzene yielded $z_0 \approx 3.13 \pm 0.09$~\AA{} on Cu, Ag, and Au (111) surfaces\cite{Bnz_geometry}.  However, the measured heights of other $\pi$-conjugated molecules can vary by as much as 0.5~\AA{} on the same substrates\cite{DFTScreeningvdWPRL}. Although such a constant height model is a clear approximation, we may reasonably expect the light--molecule/surface interaction to exhibit the same qualitative behaviour within $\pm 0.5$~\AA, as was the case for fullerene\cite{Submitted}.  For these reasons, only for weakly-interacting systems, where omni-directional van der Waals forces dominate, are we justified in describing the surface using a simple jellium model, neglecting its corrugation.

The molecule's KS orbitals $\psi_n({\bf r})$ are obtained by using the plane-wave self-consistent field DFT code \textsc{PWscf} of the \textsc{Quantum Espresso} (QE) package,\cite{QE} within the generalized gradient approximation (GGA) of Perdew and Wang (PW91) \cite{PW91-GGA} for the xc functional.  For carbon atoms we used GGA-based ultra-soft pseudopotentials,\cite{pseudopotentials} and found the energy spectrum to be converged with a $30$~Ry plane-wave cutoff.  It should be noted that the mixture of LDA and GGA xc functionals was for practical reasons, and should not impact the results of our $G_0W_0$-BSE and $G_0[W_0+\Delta W]$-BSE calculations.

To describe the molecule--surface interaction, we have used the quasiparticle $G_0W_0$ eigenvalues calculated with \textsc{vasp} along with the KS orbitals from \textsc{PWscf} and a somewhat reduced number of levels $N_L$. For the determination of the screened interaction $W$, which enters into the $\textit{G}_{\mathrm{0}}\textit{W}_{\mathrm{0}}$ scheme, we included transitions between $15$ occupied and $55$ unoccupied, $33$ occupied and $147$ unoccupied, and $120$ occupied and $180$ 
unoccupied molecular orbitals for benzene, terrylene and fullerene respectively, as shown in Table~\ref{Table1}. 

To model a single isolated molecule, we must exclude the effect on its polarizability due to the interaction with the surrounding molecules in the lattice. This is accomplished by solving the BSE using a truncated Coulomb interaction \cite{RadialCutoff}
\begin{equation} 
V_C({\bf r}-{\bf r}')=\frac{\Theta\left(R_C - |{\bf r}-{\bf r}'|\right)}{|{\bf r}-{\bf r}'|},
\label{RadialCutoff}
\end{equation}
where $\Theta$ is the Heaviside step function, and $R_C$ is the range of the Coulomb interactions, i.e., the radial cutoff. Since the lattice constant $L$ is more than twice the range of the fullerene molecule's density, using a radial cutoff of $R_C=L/2$ ensures that the charge fluctuations created within the molecule produce a field throughout the whole molecule, but do not produce any field within the surrounding molecules. An important property of the truncated Coulomb interaction $V_C$ is that it remains invariant upon translation in reciprocal space.  However, it becomes nonlocal when the surface is introduced, as it then depends explicitly on the height above the surface. 

In order to include the influence of surface screening, the ${G_\mathrm{0}}\textit{W}_{\mathrm{0}}$ quasiparticle energies are additionally  corrected by using ${G_\mathrm{0}}[W_0+\Delta\textit{W}]$, a method proposed in refs~\citenum{G0W0DeltaW2010,Submitted,Bnz_absorption_theory2,Spataru,G0W0DeltaW2017}\nocite{G0W0DeltaW2010,Submitted,Bnz_absorption_theory2,Spataru,G0W0DeltaW2017}. In this method $\Delta\textit{W}$ represents the surface induced dynamical Coulomb interaction, and the molecular optical absorption spectra is calculated by solving the BSE. The influence of the metal surface is included such that the bare Coulomb interaction $V$ is replaced by the screened Coulomb interaction $\textit{V}+\Delta\textit{W}$ throughout, and the screening $W$ in BSE is replaced by $W_0+\Delta W$\cite{Bnz_absorption_theory2}. In this way surface screening is included implicitly in the exchange term\cite{ExchangeScreeningThygesenPRB2019} and explicitly in the correlation term of the self energy $\Sigma$ \cite{Bnz_absorption_theory2}. This yields an effective $G_0[W_0+\Delta W]$-BSE method \cite{Bnz_absorption_theory2,Submitted}.

The benzene bright exciton \Eu{} is mostly composed of transitions between benzene's occupied $\{\sigma,\pi_2,\pi_3\}$ and unoccupied $\{\pi_4^*,\pi_5^*,\sigma^*\}$ orbitals, where both $\{\pi_2,\pi_3\}$ and $\{\pi_4^*,\pi_5^*\}$ are degenerate\cite{Bnz_absorption_theory2}. Since the \Eu{} exciton is the main feature of benzene's absorption spectra\cite{Bnz_absorption_exp3}, we may restrict consideration to transitions within $\{$HOMO$-2,$ HOMO$-1,$ HOMO$,$ LUMO$,$ LUMO$+1,$ LUMO$+2\}$\cite{Bnz_absorption_theory2} in our $G_0[W_0+\Delta W]$-BSE calculations, as stated in Table~\ref{Table1}. 

The terrylene bright \Bu{} exciton corresponds to transitions between the HOMO and the LUMO.  For this reason, we only need to include the HOMO and LUMO levels when inverting the BSE kernel for terrylene, as stated in Table~\ref{Table1}. 

To describe the three lowest-lying fullerene bright excitons, it is sufficient to include transitions between 14 occupied and 14 unoccupied levels\cite{Submitted}. This means when solving the BSE for fullerene we only need to include $\{$HOMO$-13,$ HOMO$-12, \ldots,$ HOMO$,$ LUMO$, \ldots,$ LUMO$+12,$ LUMO$+13\}$ \cite{Submitted}, as stated in Table~\ref{Table1}.

It is important to note that the metal surfaces we have modelled are perfect reflectors in the energy range considered herein ($0 < \hbar\omega <  9$~eV) since $\omega < \omega_{s}$.  This means the effect of the surface on the absorbance will simply be a renormalization, as the reflected light may also be absorbed by the molecule.  For this reason we may safely restrict consideration to the optical absorbance of the molecules themselves.

\section{RESULTS AND DISCUSSION}

\subsection{Metal Substrate Model}

To obtain a complete understanding of how surface excitations effect the optical absorption properties of the molecule, we begin by considering the energy alignment of the two systems in isolation: the molecule in gas phase and a clean metal substrate.  

\begin{figure}[!htb]
  \centering
\includegraphics[width=0.9\columnwidth]{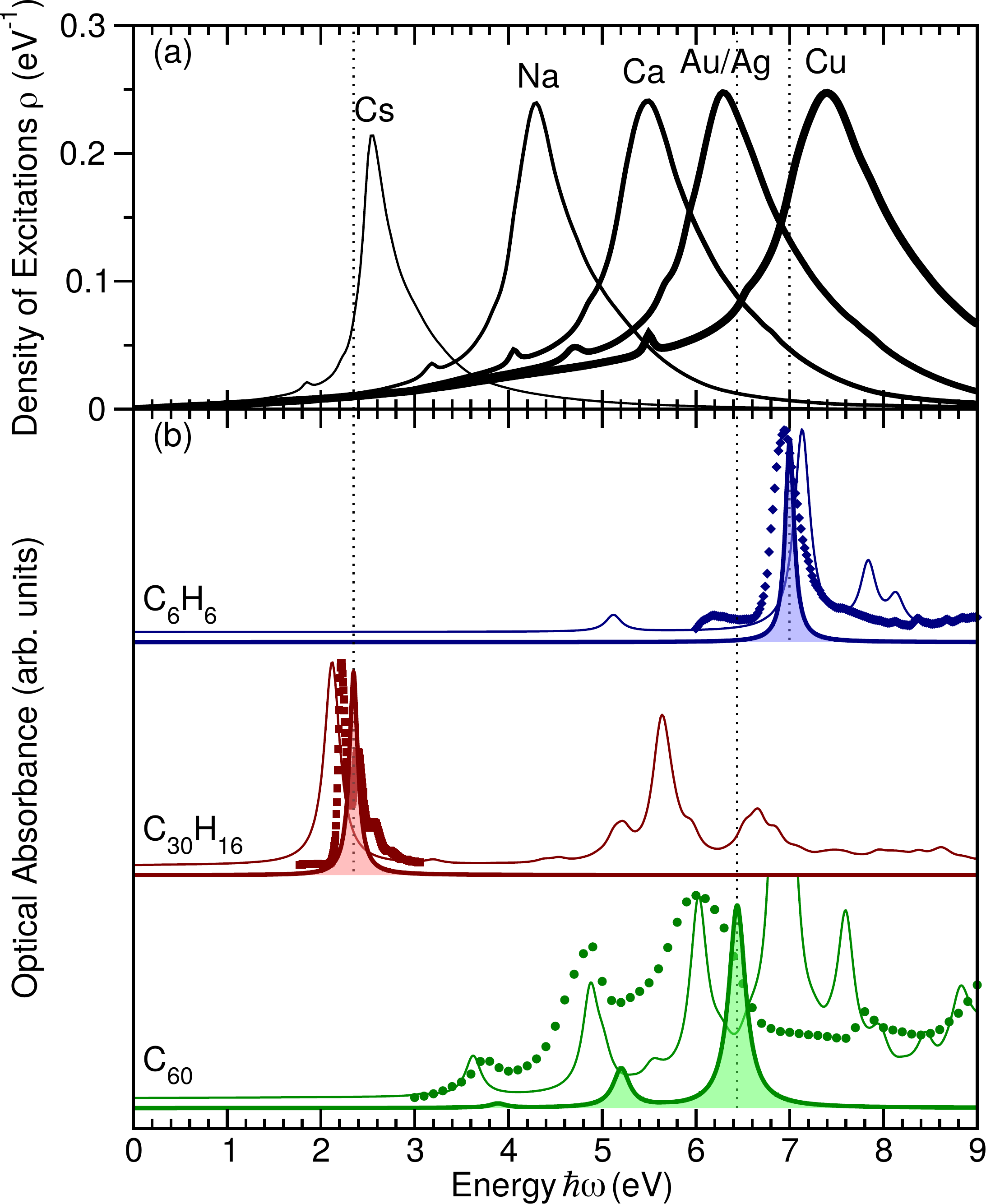}
\caption{(a) Density of electron-hole and collective (surface plasmon) excitations $\rho$ in eV$^{-1}$ for various metal surfaces\cite{Leo}. (b) Measured (symbols) and $G_0W_0$-BSE calculated (lines) optical absorbance of an isolated benzene (blue, diamonds\cite{Bnz_absorption_exp3}), terrylene (red, squares\cite{TerryleneAbsorbance}), and fullerene (green, circles\cite{Fullerene-opt2}) molecule versus energy $\hbar\omega$ in eV.  The number of occupied and unoccupied levels included are those employed in $G_0W_0$-BSE (thin lines) and $G_0[W_0+\Delta W]$-BSE (thick lines) calculations, as provided in Table~\ref{Table1}.  The molecules' exciton frequencies $\omega_{\textit{ex}}$ (dotted black lines) are shown to guide the eye.}
\noindent{\color{StyleColor}{\rule{\columnwidth}{1pt}}}
\label{Fig2} 
\end{figure}

Figure~\ref{Fig2}(a) shows the alignment of the metal substrate's density of excitations $\rho$, i.e., electron-hole and collective surface plasmon excitations
\begin{equation}
  \rho(\omega) = -\frac{1}{2\pi^2}\Imag\left[\int_0^\infty \frac{1-\varepsilon(Q,\omega)}{1+\varepsilon(Q,\omega)} Q dQ\right].\label{excitationdensity}
\end{equation}
Here, integration is over the magnitude $Q$ of the 2D wavevector $\Qv = \{Q_x,Q_y\}$ in the surface plane. The surface dielectric function  $\varepsilon(Q,\omega)$ is calculated within RPA including local crystal field effects (LCF) normal to the surface with the KS orbitals of the metal substrate obtained from a jellium model\cite{Leo}. These are determined self-consistently using a fixed ionic background charge density for the metal substrate of $\rho_+ = \frac{3}{4\pi} r_s^{-3} \Theta(-z)$~Ha$^{-1}$, where $z$ is the height above the substrate and $r_s$ is the metal's Wigner-Seitz radius in $a_0$ from ref~\citenum{AshcroftMermin}\nocite{AshcroftMermin}.

Figure~\ref{Fig2}(a) shows that as $r_s$ increases and $\rho_+$ decreases from Cu to Cs, the main peak in the density of electron-hole and collective excitations, i.e., the surface plasmon frequency, is shifted to lower energy.  In this way, by appropriately choosing the Wigner-Seitz radius of the metal substrate, we can tailor the surface plasmon energy, $\hbar\omega_s = \sqrt{\text{\sfrac{3}{2}}} r_s^{-\text{\sfrac{3}{2}}}$~Ha, to align with the bright excitons $\omega_{\textit{ex}}$ of the molecules.

It is important to note that this jellium model for the surface's dielectric function contains not only surface plasmons but also single-particle electron-hole excitations. In this way, when a molecule is included in this surface model, its excitons not only hybridize with surface plasmons but also interact with the continuum of surface electron-hole excitations, potentially causing their irreversible decay.  Although this jellium model neglects interband transitions, which could prove important for Au and Cu surfaces, we expect these effects will be dominated by both the intraband ``Drude'' tail's contribution to the molecular decay rate and the surface plasmon's screening of the interband electron-hole excitations.  Moreover, even if the energy of the surface plamon is altered by the inclusion of interband electron-hole excitations, it will remain the dominant resonance in electron energy loss spectroscopy (EELS).  This justifies our use of the surface plasmon to modify the molecular exciton's energy.

\subsection{Gas-Phase Molecular Absorbance}
The optical absorption spectra for the molecules in gas phase from $G_0W_0$-BSE calculations are shown in Figure~\ref{Fig2}(b) for benzene, terrylene, and fullerene. These spectra consist of a series of narrow peaks at the molecule's exciton energies $\hbar\omega_{\textit{ex}}$.  To facilitate a direct comparison between theory and experiment, in Table~\ref{Table2} we tabulate the energies $\hbar\omega_{\textit{ex}}$ for the bright excitons of benzene, terrylene, and fullerene.
\begin{table}
\caption{\normalsize{\textrm{\bf Bright Exciton Energies $\bm{\hbar\omega}_{\textit{ex}}$ in Electronvolts from \emph{G}$_{\text{0}}$\emph{W}$_{\text{0}}$-BSE, Number of Levels Included \emph{N}$_{\textit{L}}$, and Optical Absorbance Spectroscopy (OAS) and Electron Energy Loss Spectroscopy (EELS) Measurements for Benzene (C$_{\text{6}}$H$_{\text{6}}$), Terrylene (C$_{\text{30}}$H$_{\text{16}}$), and Fullerene (C$_{\text{60}}$)}}}\label{Table2}

  \begin{tabular}{lcr@{\quad}lr@{\quad}l@{}r@{}c@{}}
    \multicolumn{8}{>{\columncolor[gray]{.9}}c@{}}{}\\[-1.0em]    
    \multicolumn{1}{>{\columncolor[gray]{.9}}c@{}}{molecule} &
    \multicolumn{1}{>{\columncolor[gray]{.9}}c@{}}{exciton} &
    \multicolumn{1}{>{\columncolor[gray]{.9}}c}{\emph{N}$_{\textit{L}}$} &    
    \multicolumn{1}{>{\columncolor[gray]{.9}}l}{$\!\!\!\!\!\!\hbar\omega_{\textit{ex}}$} &
    \multicolumn{1}{>{\columncolor[gray]{.9}}c}{\emph{N}$_{\textit{L}}$} &    
    \multicolumn{1}{>{\columncolor[gray]{.9}}l}{$\!\!\!\!\!\!\hbar\omega_{\textit{ex}}$} &
    \multicolumn{1}{>{\columncolor[gray]{.9}}c@{}}{OAS} &
    \multicolumn{1}{>{\columncolor[gray]{.9}}l@{}}{EELS} \\[0.5mm]
    C$_{\text{6}}$H$_{\text{6}}$ & \Eu{} & 60 & 7.13 & 6 & 7.00 & 6.94$^a$\\
    C$_{\text{30}}$H$_{\text{16}}$ & \Bu{} & 188 & 2.12 & 2 & 2.35 & 2.22$^b$ \\   
    C$_{60}$ & $\pi-\pi^*$ & 240 & 3.62 & 28 & 3.89 & 3.75$^c$ & 3.7$^d$ \\
    C$_{60}$ & $\pi-\pi^*$ & 240 & 4.88 & 28 & 5.21 & 4.90$^c$ & 4.8$^d$\\
    C$_{60}$ & $\pi-\pi^*$ & 240 & 6.02 & 28 & 6.44 & 6.00$^c$ & 6.3$^d$\\
    \multicolumn{8}{p{0.85\columnwidth}}{\footnotesize{$^a$Reference~\citenum{Bnz_absorption_exp3}. $^b$Reference~\citenum{TerryleneAbsorbance}. $^c$Reference~\citenum{Fullerene-opt2}. $^d$Reference~\citenum{Lucas2}.\nocite{Bnz_absorption_exp3,TerryleneAbsorbance,Fullerene-opt2,Lucas2}}}
  \end{tabular}

\noindent{\color{StyleColor}{\rule{\columnwidth}{1pt}}}
\end{table}

\begin{figure}[!htb]
  \includegraphics[width=0.9\columnwidth]{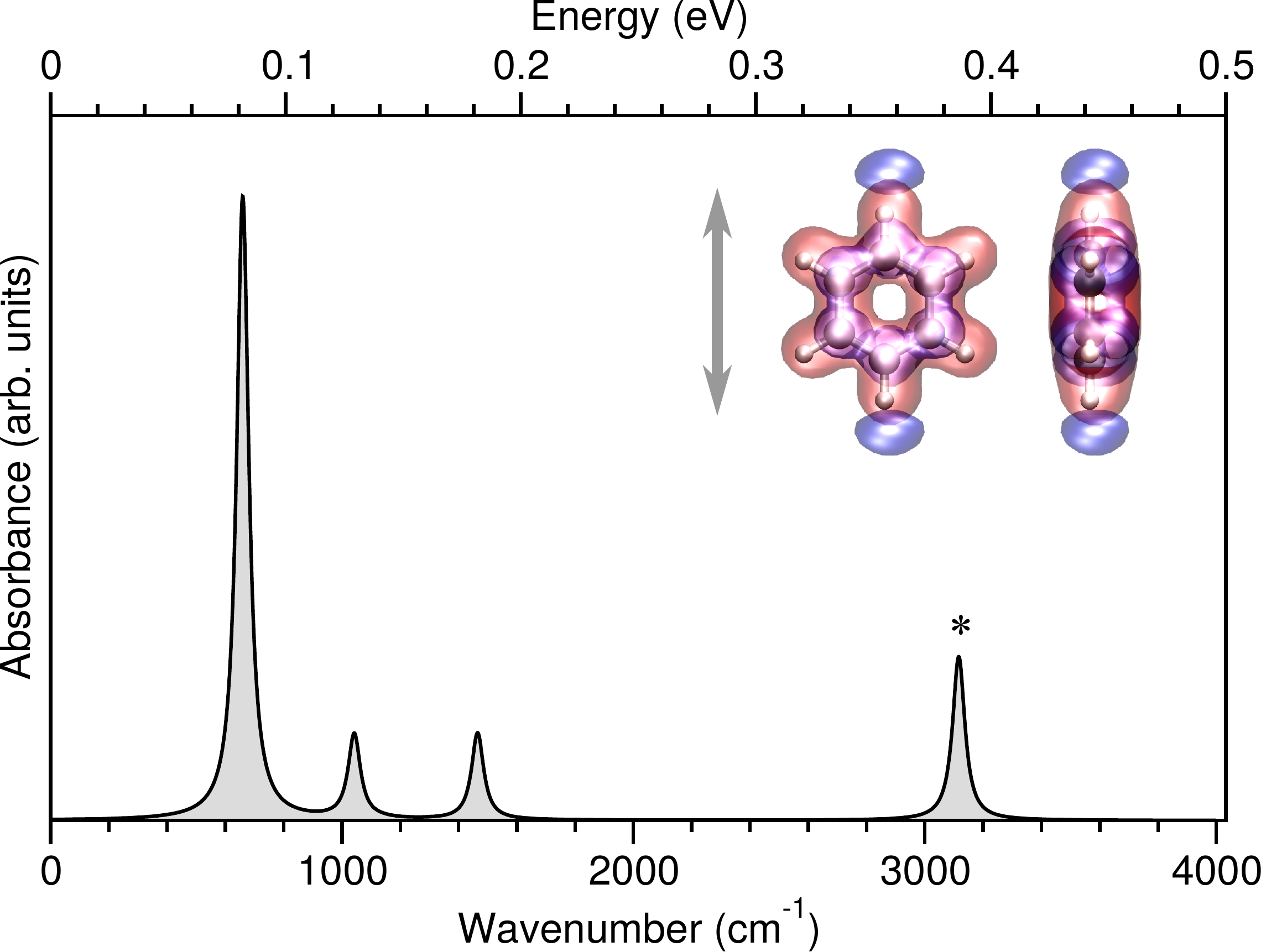}
  \caption{Fourier transformed infrared (FTIR) spectra of benzene in gas phase.  Inset shows $G_0W_0$-BSE excited electron (blue) and hole (red) densities of benzene's bright \Eu{} exciton at 7.1~eV, with polarization depicted by arrows. This exciton could couple to the vibrational mode (*) at 0.39~eV shown in Supporting Information.}\label{BenzeneFTIR}
    \noindent{\color{StyleColor}{\rule{\columnwidth}{1pt}}}
\end{figure}

For benzene (C$_6$H$_6$), the first bright peak at $\hbar\omega_{\textit{ex}} \approx 7.0$~eV  (\emph{cf.}\ Figure~\ref{Fig2}(b) and Table~\ref{Table2}), corresponds to its \Eu{} exciton.  This exciton's electron ($\rho_e(\rv_e)$) and hole ($\rho_h(\rv_h)$) exciton densities from our gas-phase $G_0W_0$-BSE calculations are shown in the inset of Figure~\ref{BenzeneFTIR}.
From these spatial distributions it is clear that the \Eu{} exciton moves the electron from a bonding $\sigma$ to an antibonding $\sigma^*$ orbital of the C--H bonds aligned with the polarization axis.  Such an excitation could potentially promote the dehydrogenation of the benzene ring if coupled to an appropriate vibrational mode.

Our calculated FTIR spectrum (\emph{cf.}\ Figure~\ref{BenzeneFTIR})  shows that the C--H stretching vibrational mode shown in the Supporting Information (SI) at 0.39~eV is IR active, and could couple with the \Eu{} exciton.  From a careful examination of benzene's measured optical absorbance\cite{Bnz_absorption_exp3} shown in Figure~\ref{Fig2}(b), we see evidence of satellite peaks at 7.4 and 7.8~eV.  These peaks could be due to electron-phonon coupling between the electronic states of the \Eu{} excitation and the C--H stretching mode.  However, calculations to verify this and quantify their electron-phonon coupling explicitly are beyond the scope of this work. Altogether, these results suggest that, with the aid of a sufficiently active substrate (e.g.\ Cu(111)\cite{CatalysisRev}), the photocatalytic dehydrogenation of benzene may be induced by ultraviolet (UV) light ($\lambda \approx 168$~nm) exciting both an electron to a C--H antibonding orbital and the C--H stretching vibrational mode.

\begin{figure}[!ht]
  \includegraphics[width=0.9\columnwidth]{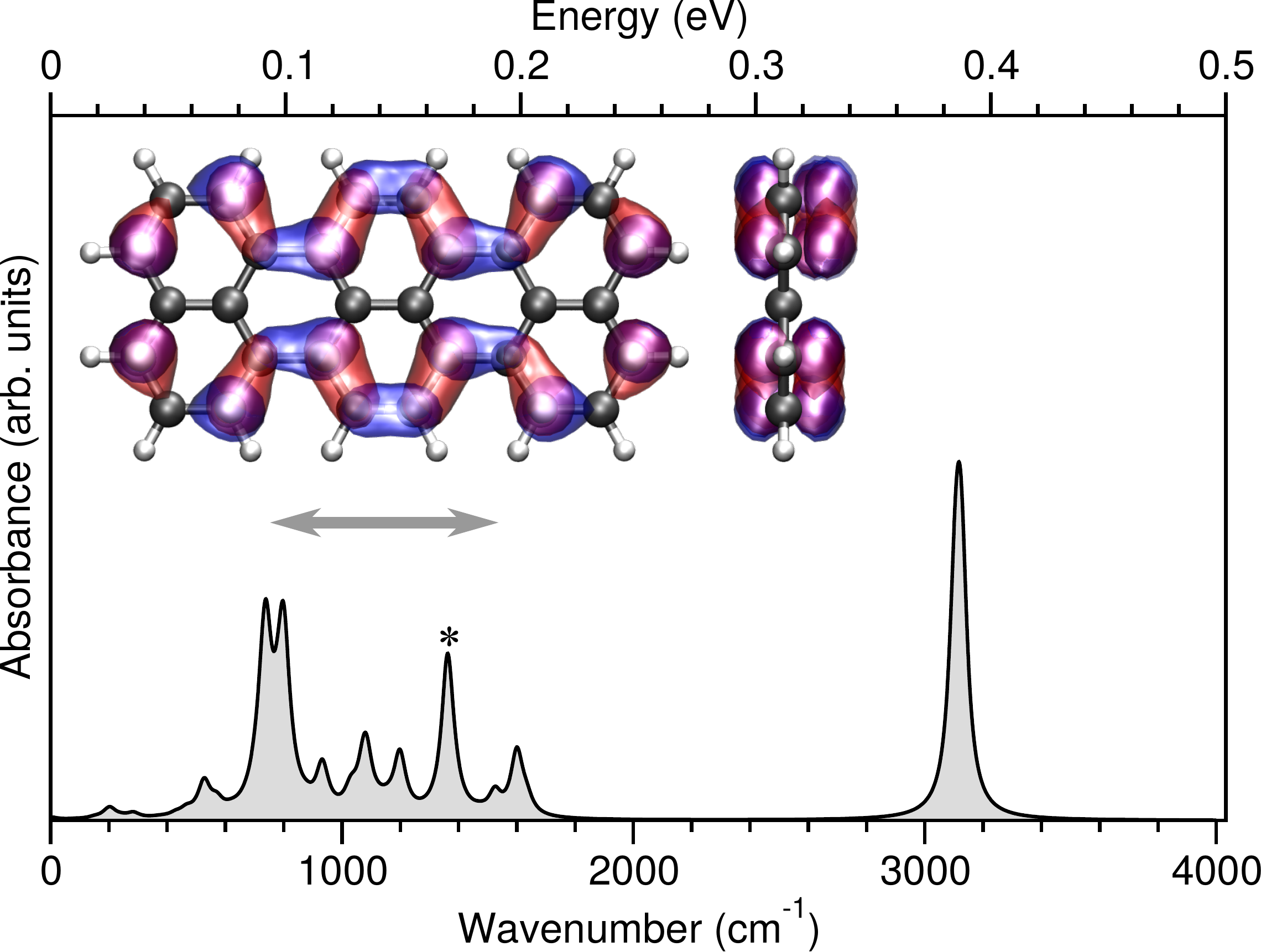}
  \caption{Fourier transformed infrared (FTIR) spectra of terrylene in gas phase.  Inset shows $G_0W_0$-BSE excited electron (blue) and hole (red) densities of terrylene's bright \Bu{} exciton at 2.1~eV, with polarization depicted by arrows. This exciton could coupleold{s} to the vibrational mode (*) at 0.17~eV shown in Supporting Information.}\label{TerryleneFTIR}
  \noindent{\color{StyleColor}{\rule{\columnwidth}{1pt}}}
\end{figure}

For terrylene (C$_{30}$H$_{16}$), the first bright peak at $\hbar\omega_{\textit{ex}} \approx 2.2$~eV (\emph{cf.}\ Figure~\ref{Fig2}(b) and Table~\ref{Table2}), corresponds to the \Bu{} exciton.  This exciton's electron ($\rho_e(\rv_e)$) and hole ($\rho_h(\rv_h)$) exciton densities from our gas-phase $G_0W_0$-BSE calculations are shown in the inset of Figure~\ref{TerryleneFTIR}.
From these spatial distributions we see that the \Bu{} exciton moves the electron from a bonding $\pi$ to an antibonding $\pi^*$ orbital of the outer phenyl rings, resulting in their twisting.  Such an excitation could potentially lead to a decomposition of terrylene's outer phenyl rings if coupled to an appropriate vibrational mode.

From our calculated FTIR spectrum  (\emph{cf.}\ Figure~\ref{TerryleneFTIR}) we find a twisting vibrational mode of the outer phenyl rings shown in the SI at 0.17~eV is IR active, and could couple with terrylene's \Bu{} exciton.  From the measured high resolution optical absorption spectra of terrylene\cite{TerryleneAbsorbance} shown in Figure~\ref{Fig2}(b), we may already clearly distinguish two satellite peaks at 0.17 and 0.34~eV above terrylene's bright \Bu{} exciton.  This quantitative agreement with the calculated vibrational energies suggests that these satellite peaks could be due to electron-phonon coupling between the electronic states of the \Bu{} excitation and the twisting vibrational mode of terrylene's outer phenyl rings.  However, explicit calculations of their electron-phonon coupling to quantify this are again beyond the scope of this work.

Having understood the nature of benzene's \Eu{} and terrylene's \Bu{} excitons and their coupling to vibrational modes of the molecules, in an effort to tailor the energies and lifetimes of these excitations, we may employ hybridization between these excitons and the surface plasmons of the substrate.  Qualitatively, the degree of interaction and/or hybridization between the molecule's excitons and the metal substrate's surface plasmons will be determined by the energy alignment between $\omega_{\textit{ex}}$ and $\omega_s$.  Since pure transition metals have Wigner-Seitz radii spread throughout the range $1.8\leq r_s \leq 5.6~a_0$, i.e., $2.5 \leq \hbar \omega_s \leq 10$~eV, by alloying these metals  one should always be able to find a substrate with $\omega_s \approx \omega_{\textit{ex}}$.  In fact, the $\omega_s \approx \omega_{\textit{ex}}$ condition is reasonably satisfied by Cu for benzene's (C$_6$H$_6$) first bright \Eu{} exciton; Cs for terrylene's (C$_{30}$H$_{16}$) first bright \Bu{} exciton; and Na, Ca, and Au/Ag for fullerene's (C$_{60}$) first three bright $\pi-\pi^*$ excitons, as shown in Figure~\ref{Fig2}.  With these results in mind, we shall consider the interfacial light--molecule interaction over the  complete $1.5 \leq r_s \leq 7~a_0$ range in the following sections. 
\subsection{Physisorbed Benzene Optical Absorption}

\begin{figure}[!htb]
\includegraphics[width=\columnwidth]{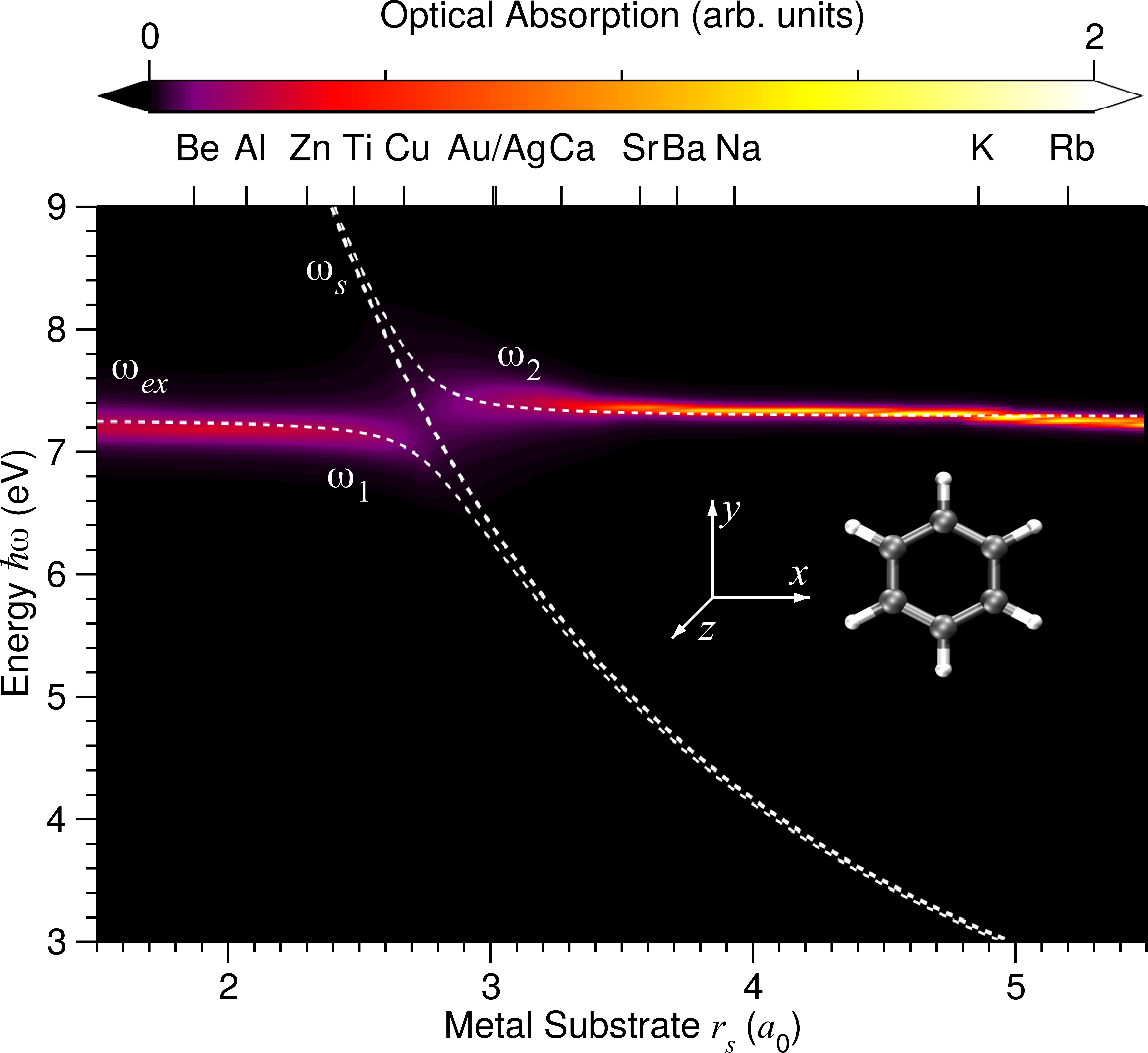}
\caption{Benzene's $G_0[W_0+\Delta W]$-BSE optical absorption intensity for physisorbed benzene as a function of the incident $x$-polarized light's energy $\hbar\omega$ in eV and the metal substrate's Wigner-Seitz radius $r_s$ in $a_0$  at a height $z_0=6~a_0 \approx 3.18$~\AA{} above the substrate. $G_0W_0$-BSE excited electron and hole densities for the bright \Eu{} exciton in gas phase are shown in Figure~\ref{BenzeneFTIR}.  The exciton frequency $\omega_{\textit{ex}}$, surface plasmon frequency $\omega_s$, and their hybridized modes $\omega_{1,2}$ from (\ref{omega12}) are marked.}
\noindent{\color{StyleColor}{\rule{\columnwidth}{1pt}}}
\label{Fig1} 
\end{figure}

Figure~\ref{Fig1}
shows the $G_0[W_0+\Delta W]$-BSE optical absorption intensity as a function of the metallic substrate's 
Wigner-Seitz radius $r_s$. The incident light is $x$-polarized and the molecule and surface lie in the $x y$-plane, as shown in the inset of Figure~\ref{Fig1}, with the molecule $z_0= 6~a_0 \approx 3.18$~\AA{} above the substrate. This corresponds to the average most stable height for benzene above Cu, Ag, and Au (111) of $z_0 \approx 3.13$~\AA{} from PBE-dvW calculations\cite{Bnz_geometry}. The dashed line shows the energy of the surface plasmon $\hbar\omega_s = \sqrt{\text{\sfrac{3}{2}}} r_s^{-\text{\sfrac{3}{2}}}$~Ha $\approx 33.327 r_s^{-\text{\sfrac{3}{2}}}$~eV as a function of $r_s$ in $a_0$.

Figure~\ref{Fig1} shows benzene's first bright \Eu{} exciton, whose energy is $\hbar\omega_{\textit{ex}} \approx 7.0$~eV, as obtained in many experimental and theoretical studies \cite{Bnz_absorption_exp1,Bnz_absorption_exp2,Bnz_absorption_exp3,Bnz_absorption_theory1,Bnz_absorption_theory2}. This is the lowest energy exciton in the optical spectra and can be considered as a molecular optical excitation threshold or optical gap.

The energetic hybridization between the surface plasmon $\hbar\omega_s$ and the exciton $\hbar \omega_{\textit{ex}}$ may be successfully modeled as a two-level quantum system with coupling energy $V$ and Hamiltonian $H = \left(\begin{array}{cc}\hbar \omega_s & V\\V^\dagger & \hbar \omega_{\textit{ex}}\end{array}\right)$.  The resulting hybridized eigenenergies
\begin{equation}
\omega_{1,2} = \frac{\omega_s+\omega_{\textit{ex}}}{2} \mp \sqrt{\frac{(\omega_s-\omega_{\textit{ex}})^2}{4} + \frac{|V|^2}{\hbar^2}}\label{omega12}
\end{equation}
may then be fitted to the peaks in the optical absorption spectra using the coupling energy $V \approx 0.36$~eV, as shown in Figure~\ref{Fig1}.

As can be seen in Figure~\ref{Fig1}, the metallic surfaces weakly affect benzene's optical gap, 
i.e., the \Eu{} exciton branch is almost horizontal. Even when the surface plasmon crosses the exciton ($\omega_{s} \approx \omega_{\textit{ex}}$) for $r_s \approx 2.8~a_0$, this does not noticeably affect the exciton's energy. This suggests that there is only a weak hybridization between the \Eu{} exciton and the surface plasmon, which is attributable to benzene's smaller area relative to terrylene and fullerene.

However, we can see that for $r_s\lesssim 3.3~a_0$ the exciton lifetime is quite short, while the opposite is true for $r_s\gg 3.3~a_0$, where it becomes quite long lived. Since for $r_s \gg 3.3~a_0$ the exciton's width is less than the broadening $\eta\approx 20$~meV employed in the calculations, the \Eu{} exciton is basically uncoupled from the substrate.  Surprisingly\cite{C60quenching,ExcitonQuenching,QuenchingMetalNPs}, this suggests benzene's \Eu{} exciton would not be quenched by adsorption on transition metal substrates with $r_s \gg 3.3~a_0$, e.g., Na, K, and Rb.  Altogether, this means one may use the substrate's Wigner-Seitz radius to tailor the lifetime of physisorbed benzene's \Eu{} exciton.

\subsection{Physisorbed Terrylene Optical Absorption}

\begin{figure}[!htb]
\includegraphics[width=\columnwidth]{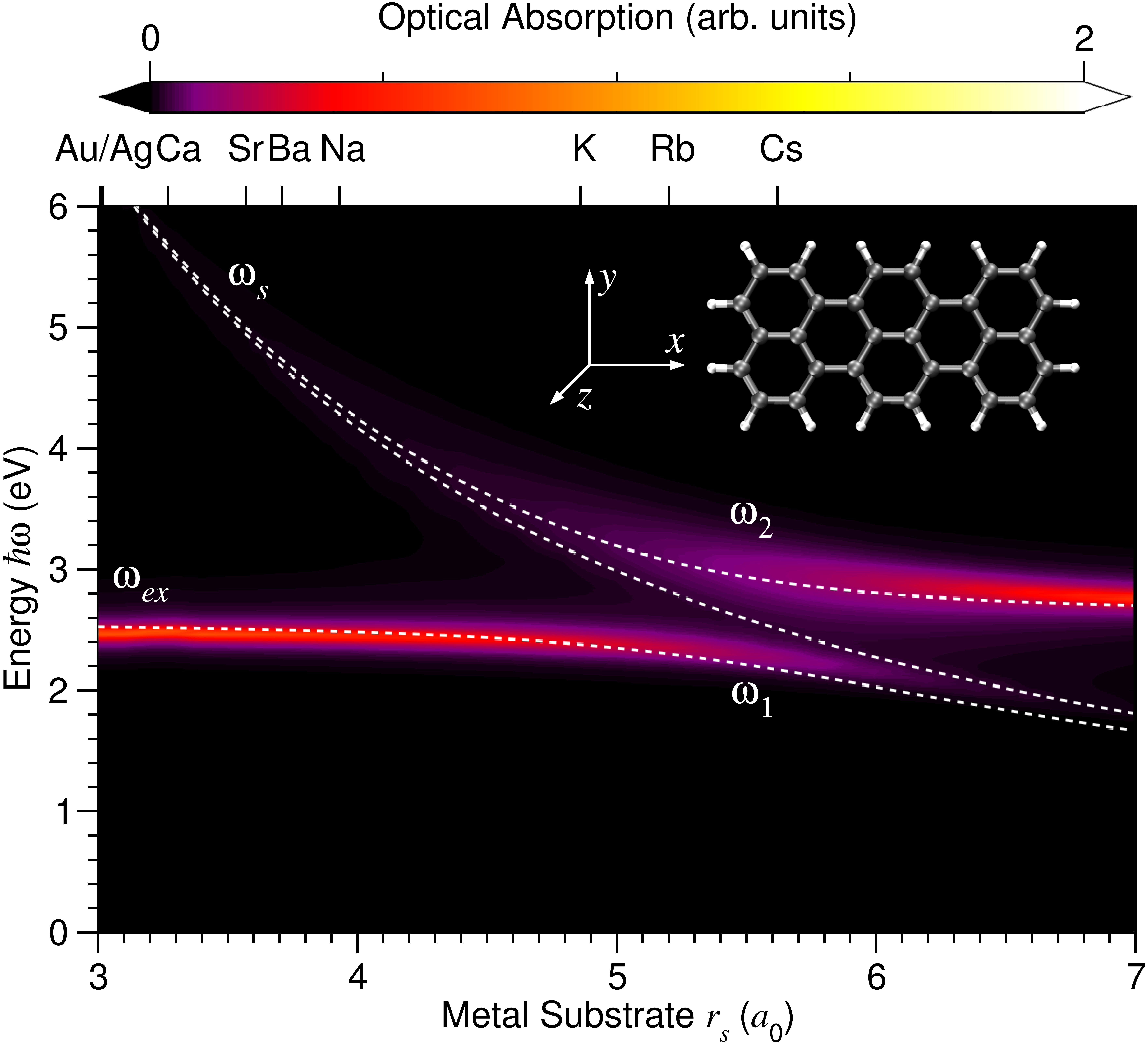}
\caption{Terrylene's $G_0[W_0+\Delta W]$-BSE optical absorption intensity as a function of the incident $x$-polarized light's energy $\hbar\omega$ in eV and the substrate Wigner-Seitz radius $r_s$ in $a_0$ at a height $z_0=6~a_0 \approx 3.18$~\AA{} above the substrate.
$G_0W_0$-BSE excited electron and hole densities for the bright \Bu{} exciton in gas phase are shown in Figure~\ref{TerryleneFTIR}.  The exciton frequency $\omega_{\textit{ex}}$, surface plasmon frequency $\omega_s$, and their hybridized modes $\omega_{1,2}$ from (\ref{omega12}) are labelled.}
\label{Fig3} 
\noindent{\color{StyleColor}{\rule{\columnwidth}{1pt}}}
\end{figure}

Figure~\ref{Fig3}
shows the $G_0[W_0+\Delta W]$-BSE optical absorption intensity for physisorbed terrylene as a function of the metallic substrate's Wigner-Seitz radius.  The incident light is $x$-polarized and the molecule and surface lie in the $x y$-plane, as shown in the inset of Figure~\ref{Fig3}, with the molecule $z_0= 6~a_0 \approx 3.18$~\AA{} above the substrate.  The dashed line represents the energy of the surface plasmon $\hbar \omega_s$ as a function of $r_s$.

Optical absorption measurements show that the isolated molecule has a strong bright exciton at $\hbar\omega_{\textit{ex}}\approx2.3$~eV\cite{Terrylene1,Terrylene2}.
As is clearly visible from Figure~\ref{Fig3}, in the case of terrylene, the surface plasmon significantly affects the molecule's absorption spectrum. We find that in 
the region $r_s\lesssim5~a_0$, apart from terrylene's bright excitation branch at $2.3$~eV, there appears an extra absorption branch at the surface plasmon 
frequency $\omega_s$. For $5 < r_s < 6~a_0$, corresponding to K, Rb, or Cs surfaces, the surface plasmon branch begins to deviate from $\omega_s$ while the exciton branch moves to lower energies. This region is of particular interest since the molecule begins to support two equally intensive bright excitons, denoted by $\omega_1$ and $\omega_2$. For $r_s\gtrsim 6~a_0$ there is an avoided crossing behaviour, where the surface plasmon and molecular exciton exchange their properties.  The upper $\omega_2$ branch 
continues to disperse as the molecular exciton, while the lower $\omega_1$ branch follows the surface plasmon branch.

The physical explanation for this peculiar behaviour is as follows\cite{Submitted}.
The incident electromagnetic field is unable to affect the metallic surface directly. Namely, incident $x$-polarized light with 
frequency below that of the surface plasmon ($\omega<\omega_s$) will be (excluding ohmic losses) completely reflected, whereas for frequency above the surface plasmon frequency ($\omega>\omega_{s}$), light will be completely transmitted. However, when light interacts with the molecule it induces dipole active charge density oscillations, i.e., excitons. Such charge oscillations produce a relatively strong and long range electric field around the molecule. If the surface is nearby, the molecule behaves as an oscillating dipole that can induce charge density oscillations in the metal surface. Moreover, the metal surface supports self-sustaining charge density oscillations, i.e., surface plasmons, that can be excited by a longitudinal probe such as a dipole.

Induced charge density waves corresponding to these modes are dispersionless plane waves 
\begin{equation}
\rho_{\bf{Q}}(\brho)=\rho_{\bQ}\cos({\bQ}\cdot\brho-\omega_{\bQ}t)
\end{equation}
whose frequency is $\omega_{\bQ}=\omega_s$, where $\brho=(x,y)$  and ${\bQ}=(Q_x,Q_y)$ represent the 2D coordinate and wave vector, respectively. If an optically excited molecular dipole oscillates at a frequency $\omega\approx\omega_s$, it will be in resonance with all $\bQ$ modes, and will excite them all simultaneously. This will cause the molecular dipole to induce a linear combination of all surface $\bQ$ modes, forming an image charge on the metal surface.  In this way, the molecular dipole creates its own image charge in the substrate with which it interacts and creates coupled modes $\omega_{1,2}$ (\ref{omega12}) that are both optically active.

For mode $\omega_1$ molecular and surface dipoles oscillate in phase, whereas for mode $\omega_2$ they oscillate out of phase.  Figure~\ref{Fig3} shows that when the molecular exciton and surface plasmons are out of resonance, i.e., for $r_s\lesssim5~a_0$ and $6~a_0 \lesssim r_s$, the molecule still weakly absorbs light at the surface plasmon frequency $\omega_s$. This is because, under these conditions, light can still induce weak charge oscillations in the molecule, which resonantly pump energy into the metal surface through the molecule's image. 
For $6~a_0\lesssim r_s$ the lower absorption branch becomes visibly shifted below $\omega_{\textit{ex}}$.  This means the surface plasmon can effectively reduce the molecule's optical gap.

\subsection{Excitonic Inverse Lifetime}

In ref~\citenum{Bnz_absorption_theory2}\nocite{Bnz_absorption_theory2} we have analyzed the interaction of benzene's excitons (triplet, singlet, bright, dark) 
with the metallic surface for $r_s=3~a_0$. We found that the width of the dark excitons $B^1_{2u}$ and $E^1_{2g}$ is completely unaffected by the metallic surface. This is unsurprising since the quadrupolar nature of these excitons ensures they are not coupled to the substrate like dipolar modes.
However, the bright \Eu{} exciton prefers to recombine such that it excites electronic modes in the metal substrate. 

On the one hand, in the metal there exist a large number or continuum of electron-hole excited states at the same energy which can be described by the density of excitations $\rho(\omega)$ (\ref{excitationdensity}), or the number of excited states per unit energy. On the other hand, the exciton is a localized or discrete excitation whose density of excitations is a Lorentzian broadened delta function $\delta(\omega-\omega_{\textit{ex}})$. If the \Eu{} exciton is immersed in such an electron-hole continuum, the Coulomb interaction will produce transitions between the \Eu{} exciton and the electron-hole excitations in the metal substrate. This causes the exciton to irreversibly dissipate its energy into incoherent electron-hole excitations in the substrate, concomitantly spreading its line shape. This scenario is analogous to the interaction between discrete impurity states and a continuum or wide band of states in a crystal lattice.  

The most appropriate method for solving such a problem is the Fano-Anderson model \cite{Fano}. If we assume that the density of electron-hole excitations is a smooth function around $\omega_{\textit{ex}}$, the final exciton density of states is then given by the Lorentzian 
\begin{equation}
A(\omega) \approx \frac{1}{\pi\hbar}\frac{\Gamma/2}{(\omega-\omega_{\textit{ex}})^2+(\Gamma/2)^2},
\label{lor}
\end{equation}
where $\hbar\omega_{\textit{ex}}$ is the exciton energy, $\Gamma$ is the exciton decay width, and the exciton lifetime is then given 
by $\tau \approx \hbar/\Gamma$. We fit the calculated spectra (for various $r_s$) to (\ref{lor}) in order to estimate the exciton's decay width or inverse lifetime $\Gamma$.

\begin{figure}
\includegraphics[width=0.7\columnwidth]{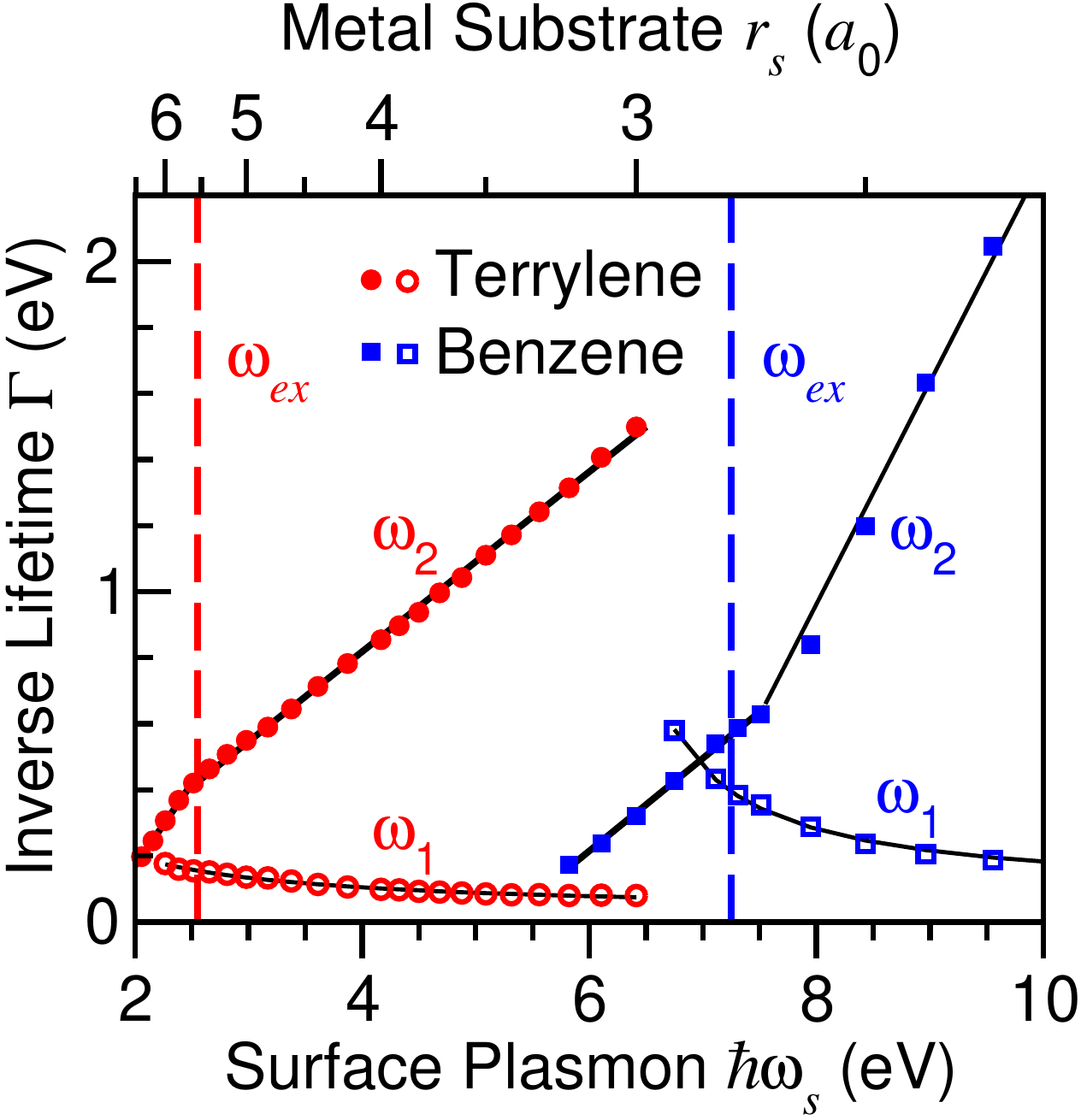}
\caption{Inverse lifetime $\Gamma$ in eV of the lower exciton $\omega_1$ (open symbols) and upper out-of-phase exciton $\omega_2$ (filled symbols) from (\ref{omega12}) for benzene (blue squares) and terrylene (red circles) versus surface plasmon energy $\hbar\omega_s$ in eV and Wigner-Seitz radius $r_s$ in $a_0$ of the metal substrate.  The exciton frequencies $\omega_{\textit{ex}}$ of benzene (short dashed blue lines) and terrylene (long dashed red lines), along with reciprocal and linear fits to $\omega_1$ and $\omega_2$, respectively, (solid black lines) are provided to guide the eye.
}\label{Lifetime}
\noindent{\color{StyleColor}{\rule{\columnwidth}{1pt}}}
\end{figure}

Figure~\ref{Lifetime}
shows how the inverse lifetime $\Gamma$ of both the lower exciton $\omega_1$ and upper out-of-phase exciton $\omega_2$ from (\ref{omega12}) for benzene and terrylene depends on the surface plasmon energy $\hbar \omega_s$ and concomitantly the Wigner-Seitz radius $r_s$, as obtained from fitting 
(\ref{lor}) to the optical absorption spectra of Figures~\ref{Fig1} and \ref{Fig3}, respectively.  For the lower exciton branch in both benzene and terrylene, the inverse lifetime $\Gamma$ depends inversely on the surface plasmon energy $\hbar\omega_s$ when $\omega_s \gtrsim \omega_{\textit{ex}}$.  If $\omega_s \gg \omega_{\textit{ex}}$, this implies $\omega_1 \sim \omega_{\textit{ex}}$ from (\ref{omega12}), and the molecular excitation becomes longer lived as the molecule decouples from the substrate. For the upper exciton branch in both benzene and terrylene, the inverse lifetime $\Gamma$ exhibits a direct linear dependence on the surface plasmon energy $\hbar\omega_s$, with a discontinuity in the derivative at $\omega_{\textit{ex}} \approx \omega_s$.  On the one hand, $\omega_s \ll \omega_{\textit{ex}}$ implies $\omega_2 \sim \omega_{\textit{ex}}$ from (\ref{omega12}). This molecular excitation becomes rather long lived as the molecule decouples from the substrate.  On the other hand, $\omega_s \gg \omega_{\textit{ex}}$ implies $\omega_2 \sim \omega_s$ from (\ref{omega12}). These surface plasmon-like excitations become increasingly shorter lived as the molecule decouples from the substrate.

Comparing these results with the characteristic molecular exciton radiative decay width of $\Gamma_{\textit{rad}}\approx 0.001$~meV in vacuum, it is clear that metallic surfaces with $\omega_s \gg \omega_{\textit{ex}}$ can cause an extraordinarily rapid decay of benzene \Eu{} and terrylene \Bu{} excitons.  These very short lifetimes reflect the expected quenching of the exciton caused by electron-hole excitations in the metal surface\cite{C60quenching,ExcitonQuenching,QuenchingMetalNPs}. Overall, the dependence of the inverse lifetime on the surface plasmon energy means we can potentially tailor the lifetime of a molecule's exciton using the Wigner-Seitz radius of the substrate.  

\subsection{Physisorbed Fullerene Optical Absorption}

\begin{figure}[!ht]
\includegraphics[width=\columnwidth]{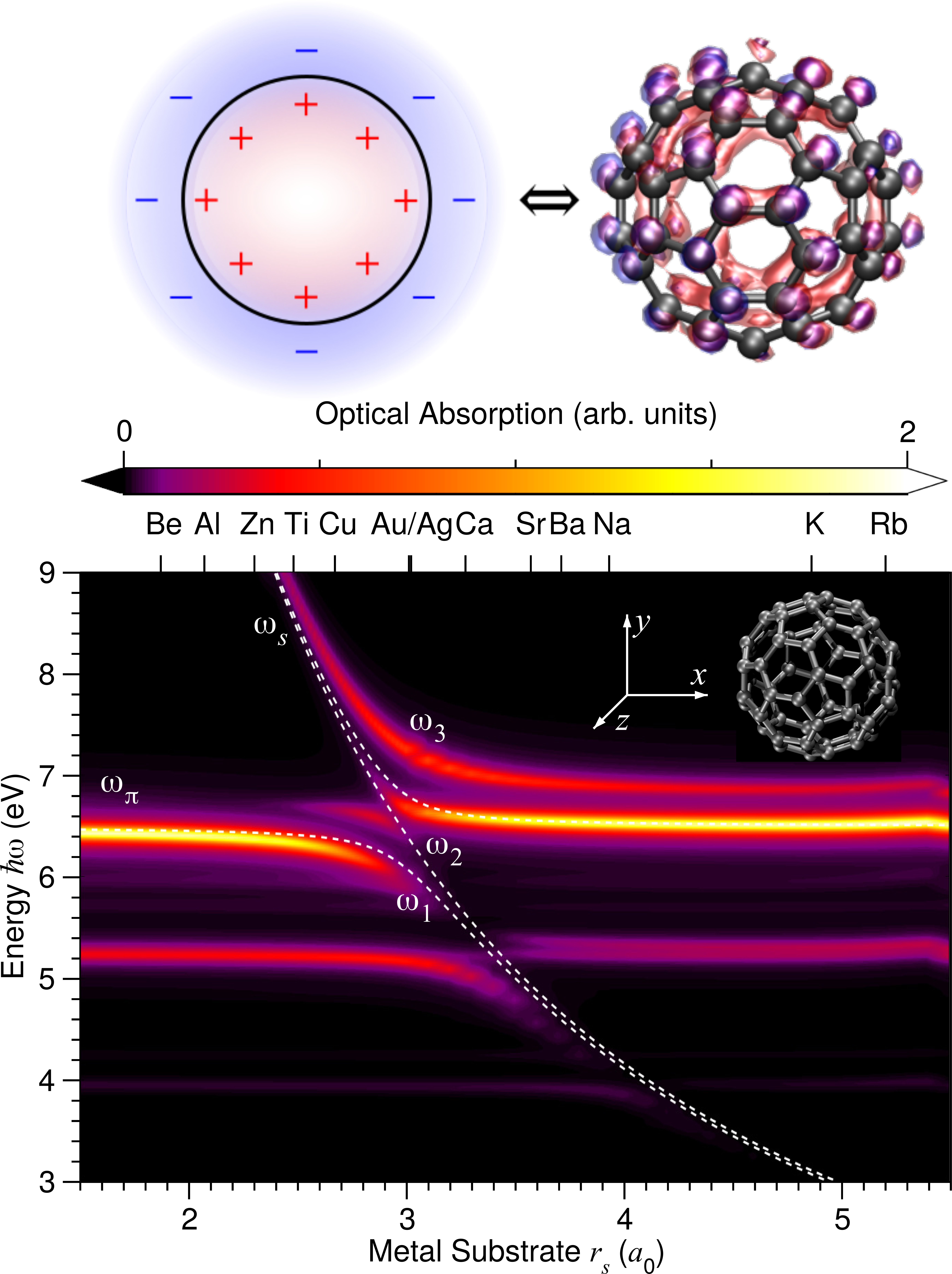}
\caption{Fullerene's $G_0[W_0+\Delta W]$-BSE optical absorption intensity as a function of the incident $x$-polarized light's energy $\hbar\omega$ in eV and the substrate Wigner-Seitz radius 
$r_s$ in $a_0$  at a minimum height $z_0=6~a_0 \approx 3.18$~\AA{} above the substrate.  The upper panel shows the excited electron (blue) and hole (red) densities schematically (left) and from $G_0W_0$-BSE calculations (right) for fullerene's third bright $\pi-\pi^*$ exciton at 6.0~eV in gas phase.  The exciton frequency $\omega_{\pi}$, surface plasmon frequency $\omega_s$, their hybridized modes $\omega_{1,2}$ from (\ref{omega12}), and the third quadrupolar exciton mode $\omega_3$ are marked.}
\label{Fig4} 
\noindent{\color{StyleColor}{\rule{\columnwidth}{1pt}}}
\end{figure}

Fullerene is a molecule whose absorption spectrum changes radically when it nears a metal surface\cite{Submitted}. Let us first consider the most intense optically active modes in isolated fullerene. As shown in Figure~\ref{Fig2}(b), the $G_0W_0$-BSE absorption spectra of isolated fullerene exhibits three main peaks. These peaks correspond to fullerene's bright excitons, which are observed experimentally at 3.8, 4.9 and 6.0~eV (\emph{cf.}\ Table~\ref{Table2}). While these energies are reproduced semi-quantitatively at the $G_0W_0$-BSE level ($\Delta\epsilon \approx 1$\%), for the reduced number of bands employed in our $G_0[W_0+\Delta W]$ calculations (\emph{cf.}\ Table~\ref{Table1}) the exciton energies are slightly blue shifted ($\Delta\epsilon \approx 5\%$) to 3.9, 5.2 and 6.4~eV, respectively (\emph{cf.}\ Figure~\ref{Fig2}(b) and Table~\ref{Table2}). The most intense peak at $\hbar\omega_{\pi} \approx 6.4$~eV corresponds to the fullerene $\pi$ plasmon resonance, which is also observed in EELS measurements at 6.3~eV\cite{Lucas2}. This is the optically active excitation whose electron and hole densities are shown schematically and from $G_0W_0$-BSE calculations as insets in Figure~\ref{Fig4}.  It is this exciton which is most affected by the metal surface.

Figure~\ref{Fig4} shows the $G_0[W_0+\Delta W]$-BSE optical absorption intensity for physisorbed fullerene as a function of the metallic substrate's Wigner-Seitz radius, where the molecule's minimum height above the surface is $z_0= 6~a_0 \approx 3.18$~\AA. The dashed line represents the energy of the surface plasmon $\hbar\omega_s$ as a function of $r_s$.

Metal surfaces with $1 \lesssim r_s \lesssim 2~a_0$, such as Be and Al, weakly affect fullerene's absorption spectra, i.e., it shows three peaks as for the isolated molecule. For $r_s \gtrsim 2~a_0$ the $\pi$ plasmon branch begins to shift to lower energies. For $r_s \gtrsim 2.5~a_0$ we find an extra higher energy branch appears in the 
optical adsorption spectra, which follows the surface plasmon energy $\omega_s$. For $r_s\approx 3~a_0$, i.e., Ag or Au surfaces, this branch is visibly shifted above $\omega_s$. At $r_s\approx 3.5~a_0$, this branch becomes a nondispersive horizontal branch, denoted by $\omega_3$. Concomitantly, the $\pi$ plasmon branch is progressively shifted to lower energies until it begins to follow the surface plasmon $\omega_s$. This lower branch is denoted by $\omega_1$. These effects are very reminiscent of an ``avoided crossing'' scenario, as was the case for terrylene. This was a consequence of the hybridization between the terrylene \Bu{} exciton and the surface plasmons.

However, for fullerene we find an unusual behaviour. Namely, the surface plasmon branch, instead of continuing as a $\pi$ plasmon, continues as an $\omega_3$ branch at slightly higher energy. At the same time, a strong nondispersive branch, denoted by  $\omega_2$, emerges at the 
$\pi$ plasmon energy. Thus, when the $\omega_\pi$ exciton is in resonance with the surface plasmon, i.e., for Ag or Au surfaces where 
$r_s\approx 3~a_0$, the interaction with the surface causes from $\omega_\pi$ to arise three separate optically active modes $\omega_1$, $\omega_2$ and $\omega_3$.

The mechanism responsible for the modes $\omega_1$ and $\omega_2$ in terrylene is easily explained using the two-level model of (\ref{omega12}). The  \Bu{} exciton $\omega_{\textit{ex}}$ interacts with its image in the metallic surface, resulting in the formation of coupled ({\em gerade} and {\em ungerade}) bright modes.  The same mechanism is responsible for forming modes $\omega_1$ and $\omega_2$ in fullerene. However, the mechanism responsible for the appearance of mode $\omega_3$ is more complicated\cite{Submitted}.

To understand this phenomenon, we must consider the symmetry of all isolated molecular transitions, i.e., far from the metallic surface, and whether they are optically active or not. A linearly polarized electromagnetic field is only able to excite modes which 
have dipolar character, whereas quadrupolar modes are not excitable. This means the quadrupolar modes of a molecule are not 
visible in optical absorption spectra.  To excite all types of modes one requires an asymmetric probe. 
One way to do this is to examine the energy loss of an oscillating dipole placed close to the molecule. 

In ref~\citenum{Submitted}\nocite{Submitted} we showed that if fullerene is excited by an $x$-polarized oscillating dipole placed 6.35~\AA{} from the molecule's 
centroid the energy loss spectrum shows two strong peaks with frequencies $\hbar\omega_+\approx 6.4$ and $\hbar\omega_-\approx6.8$~eV. 
The mode $\omega_+$ has dipolar character and corresponds to the optically active $\pi$ plasmon. In the optical absorption spectra this mode appears at frequency $\omega_\pi$. The mode $\omega_-$ has quadrupolar character and is not visible in the optical absorption spectra, i.e., it is an optically inactive mode.

As fullerene approaches an Ag or Au surface, the molecular charge density oscillations begin to interact with the surface plasmon\cite{Submitted}. 
Because the $\omega_\pm$ plasmons are in resonance with the Ag or Au surface plasmon $\omega_s$ (\emph{cf.}\ Figure~\ref{Fig4}), the interaction is strong and breaks the symmetry of the $\omega_\pm$ modes. This strong interaction with the metal surface plasmon $\omega_s$ splits the $\omega_\pm$ modes, breaks the pure quadrupole character of the $\omega_-$ mode, and  it begins to interact with the electromagnetic field, i.e., to be optically active. In this case two molecular modes $\omega_\pm$ and the surface plasmon $\omega_s$ form three coupled 
modes $\omega_1$, $\omega_2$ and $\omega_3$ which are all optically active\cite{Submitted}. 

Mode $\omega_1$ is created from the molecular symmetric mode $\omega_+$, but oscillates out of phase with the surface charge density\cite{Submitted}. Further, the induced charge is more localized 
in the upper edge of the molecule. Mode $\omega_2$ is created from the molecular symmetric mode $\omega_+$, and oscillates in phase with the surface charge density\cite{Submitted}. 
In this case the induced charge is mostly localized at the molecule--surface interface. The third mode, $\omega_3$, is created from the molecular asymmetric mode, $\omega_-$, such that charge density 
at the molecule--surface interface oscillates in phase\cite{Submitted}. 
  
So far we have considered the situation where the molecular $\pi$ plasmons $\omega_{1,2,3}$ are close to or in resonance with the surface plasmon $\omega_s$. However, for $r_s \gg 3.5~a_0$, where the surface plasmon is already much lower in  energy ($\omega_s \ll \omega_\pi$), the optical absorption spectra continues to exhibit surprising behaviour. We see that, apart from the $\pi$ plasmon which now appears as the $\omega_2$ branch, the $\omega_3$ branch continues to appear as an intensive bright mode. This means that, e.g., Na or K surfaces support one more optically active mode than the Al surface ($r_s\approx 2~a_0$).

The reason for this unusual behaviour is a consequence of the surface's response properties. 
Specifically, in a simple Drude model, the surface's response function to some external dynamical charge distribution $\rho_{\textit{ext}}(\omega)$ placed in the vicinity of the
surface is given by \cite{Pitarke}   
\begin{equation}
\chi_s(\omega)=\frac{\omega^2_s}{\omega(\omega+i\eta)-\omega^2_s},
\label{surfres}
\end{equation}
where $\eta$ is the Drude damping constant. From (\ref{surfres}) we see that if $\omega<\omega_s$ then $\chi_s(\omega)<0$. This means that the driving charge $\rho_{\textit{ext}}(\omega)$ oscillates out of phase with the induced surface charge. Otherwise, if $\omega>\omega_s$ then $\chi_s(\omega)>0$ and the induced surface charge and 
driving charge $\rho_{\textit{ext}}(\omega)$  oscillate in phase. This means surfaces with $\omega_s>\omega_\pm$, such as Al ($r_s \approx 2~a_0$), are only able to support the $\omega_1$ mode, where the molecule--surface interfacial charge densities oscillate out of phase. However, surfaces with $\omega_s<\omega_\pm$, such as Na ($r_s \approx 4~a_0$), are able to support both modes $\omega_2$ and $\omega_3$, where the molecule--surface interfacial charge densities oscillates in phase.

The metallic surface has a similar effect on the exciton located at $5$~eV. As the surface plasmon approaches, this branch shifts to lower energies. However, after crossing the surface plasmon energy, two parallel branches continue, one at 5~eV and another at slightly higher energy. The latter mode was before crossing with $\omega_s$ optically inactive. However, after crossing the surface plasmon frequency it becomes optically active.       

These effects suggest a more general conclusion. Suppose that $\omega_{\textit{ex}}$ is the frequency of a dark molecular exciton. On the one hand, if the molecule is deposited on a metallic surface with $\omega_s<\omega_{\textit{ex}}$ then the surface could convert $\omega_{\textit{ex}}$ into a bright exciton. On the other hand, metallic surfaces with $\omega_s>\omega_{\textit{ex}}$ do not affect the optical activity of the $\omega_{\textit{ex}}$ exciton.

This can potentially be applied to increase the efficiency of solar cells.   For example, molecules can be physisorbed on two different types of metallic nanoparticles with diameters $d \gg 10$~nm, so that the molecule--nanoparticle interaction is primarily on surface facets. Active nanoparticles switch on dark excitons whereas passive 
nanoparticles do not affect the optical absorption spectra. Incident light can then generate bright excitons on molecules deposited on active nanoparticles, which, via the Coulomb interaction, diffuse to molecules deposited on passive nanoparticles and become dark excitons. Such dark excitons can then no longer decay radiatively.  This should significantly increase both their lifetime and the probability of electron--hole separation.

\section{CONCLUSIONS}

Here we have studied the optical absorption and FTIR spectra of three prototypical $\pi$-conjugated molecules: benzene ($\mathrm{C}_6\mathrm{H}_6$), terrylene ($\mathrm{C}_{30}\mathrm{H}_{16}$), and fullerene ($\mathrm{C}_{60}$).  By combining gas phase $G_0W_0$-BSE and FTIR calculations, we show how electron-phonon coupling could lead to the satellite peaks in the measured optical absorption spectra for benzene and terrylene.  The spatial distribution and electron-phonon coupling of benzene's \Eu{} exciton suggests this excitation may lead to the photocatalytic dehydrogenation and polymerization of the molecule.  Similarly, the spatial distribution and electron-phonon coupling of terrylene's \Bu{} exciton suggests this excitation may lead to the photocatalytic decomposition of terrylene's outer phenyl.

We have also analyzed how adsorption on various metallic surfaces \{Be, Al, Zn, Ti, Cu, Au, Ag, Ca, Sr, Ba, Na, K, Rb, Cs\} modifies the low-lying absorption spectra of these molecules.
Benzene's bright \Eu{} exciton ($\hbar\omega_{\textit{ex}}\approx 7$~eV) weakly hybridizes with the surface plasmon, $\omega_s$. However, for $\omega_s \gg \omega_{\textit{ex}}$ this exciton quickly decays through coupling to electron-hole 
excitations in the metal substrate, e.g.,  $\Gamma \sim 1.5$~eV on Ti.   
Terrylene's bright \Bu{} exciton ($\hbar\omega_{\textit{ex}}\approx2.3$~eV) hybridizes with the substrate's surface plasmon $\omega_s$, forming
two coupled modes $\omega_{1,2}$ that are both optically active. This effect is especially strong when $\omega_s \approx \omega_{\textit{ex}}$, such as on K, Rb, and Cs surfaces. Fullerene's optically active $\pi$ exciton $\omega_+ \approx 6.4$~eV also hybridizes with the surface plasmon.  This hybridization is most efficient on Ag or Au surfaces.  Most importantly, we have demonstrated how metal surfaces can be used to convert fullerene's dark exciton ($\omega_-\approx 6.8$~eV) into an optically active bright mode when $\omega_s \lesssim \omega_-$, but not when $\omega_s \gg \omega_-$.  We anticipate these finding will motivate experimentalists to test for such behavior in not only fullerenes but also other $\pi$-conjugated molecules.  Altogether, we have shown how a proper description of interfacial light--molecular/substrate interactions is essential for the design and optimization of novel photonic materials \emph{in silico}.

 \section*{\large$\blacksquare$\normalsize{}
   ASSOCIATED CONTENT}
 \subsubsection*{\includegraphics[height=8pt]{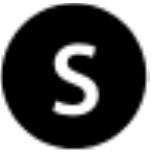}
   Supporting Information}
 \noindent
 The Supporting Information is available free of charge on the \href{http://pubs.acs.org}{ACS Publications website} at DOI: \href{http://dx.doi.org/10.1021/acs.jpcc.9b05770}{10.1021/acs.jpcc.9b05770}.\\
 
 \noindent\begin{tabular}{@{\hspace{0.075\columnwidth}}p{0.925\columnwidth}}
 Total energies and atomic coordinates of all structures and animations of relevant FTIR active vibrational modes
 \end{tabular}

\section*{\large$\blacksquare$\normalsize~AUTHOR INFORMATION}
\subsubsection*{Corresponding Authors}
\noindent *E-mail: \href{mailto:duncan.mowbray@gmail.com}{duncan.mowbray@gmail.com} (D.J.M.).\\
*E-mail: \href{mailto:vito@phy.hr}{vito@phy.hr} (V.D.).
\subsubsection*{ORCID}
\noindent Duncan John Mowbray: \href{http://orcid.org/0000-0002-8520-0364}{0000-0002-8520-0364}
\subsubsection*{Notes} 
\noindent The authors declare no competing financial interest.
\section*{\large$\blacksquare$\normalsize~ACKNOWLEDGMENTS} 
This work was supported by the QuantXLie Centre of Excellence, a project co-financed by the Croatian Government and European Union through the European Development Fund -- the Competitiveness and Cohesion Operational Program (Grant KK.01.1.1.01.0004), and used the Imbabura cluster of Yachay Tech University, which was purchased under contract No.\ 2017-024 (SIE-UITEY-007-2017).  V.D.\ is grateful to the Donostia International Physics Center (DIPC), Pedro M. Echenique, and the School of Physical Sciences and Nanotechnology, Yachay Tech University,  for their hospitality during various stages of this research. D.J.M.\ thanks the Department of Physics, University of Zagreb and Marijan \v{S}unji\'{c} for their hospitality and acknowledges funding through the Spanish ``Juan de la Cierva'' program (JCI-2010-08156), Spanish Grants (FIS2010-21282-C02-01) and (PIB2010US-00652), and ``Grupos Consolidados UPV/EHU del Gobierno Vasco'' (IT-578-13). The authors also thank I. Kup\v ci\' c for useful discussions.  


\providecommand*\mcitethebibliography{\thebibliography}
\csname @ifundefined\endcsname{endmcitethebibliography}
  {\let\endmcitethebibliography\endthebibliography}{}

\end{document}